\documentclass[aps,prb,reprint,amsmath,amssymb,superscriptaddress,floatfix,citeautoscript]{revtex4-2}
\usepackage{amsmath}
\usepackage{braket}
\usepackage{graphicx}
\usepackage{color}
\usepackage{bm}
\usepackage{braket}
\usepackage{tabularx}
\usepackage{multirow}
\usepackage[colorlinks,linkcolor=blue,citecolor=blue,urlcolor=blue]{hyperref}

\begin{document}

\title{Impact of Absorption due to Zero-Field Splitting on Loss in Dielectrics:\\A Case Study in Sapphire}

\author{Mark E. Turiansky}
\email{mturiansky@physics.ucsb.edu}
\altaffiliation[Present Address: ]{US Naval Research Laboratory, 4555 Overlook Avenue SW, Washington, DC 20375, USA}
\affiliation{Materials Department, University of California, Santa Barbara, CA 93106-5050, U.S.A.}

\author{Chris G. Van de Walle}
\email{vandewalle@mrl.ucsb.edu}
\affiliation{Materials Department, University of California, Santa Barbara, CA 93106-5050, U.S.A.}

\date{\today}

\begin{abstract}
    The coherence times of superconducting qubits are limited by loss mechanisms, whose microscopic origins have remained elusive.
    We propose a mechanism caused by transitions between zero-field-split states of paramagnetic impurities or defects.
    We derive the absorption cross section for a magnetic dipole transition and apply it to calculate the bulk loss tangent.
    For Cr, Fe, and V impurities in sapphire, we find loss tangents at 4.5 GHz in the range of $10^{-9}$--$10^{-8}$, comparable to the loss measured in experiments.
    This value suggests that magnetic loss may be a limiting factor in the coherence times of superconducting qubits.
\end{abstract}

\maketitle

\section{Introduction}
Loss in dielectric materials is a topic of high importance for many technological applications~\cite{gurevich_intrinsic_1991}.
The identification of the sources of loss has taken on new relevance in the context of quantum technologies~\cite{simmonds_decoherence_2004,martinis_decoherence_2005,oconnell_microwave_2008,de_leon_materials_2021,murray_material_2021,turiansky_dielectric_2024-1}, where loss tangents as low as 10 parts per billion are known to affect state-of-the-art superconducting qubit lifetimes~\cite{read_precision_2023}.
Dielectric loss is often attributed to two-level systems (TLSs), a model first introduced in the study of amorphous materials~\cite{thorpe_properties_2001} in which charge tunnels between two inequivalent configurations.
The coherence times of superconducting qubits improved as amorphous dielectrics were dropped in favor of their crystalline counterparts~\cite{martinis_ucsb_2014}, and surfaces or interfaces became the subject of research efforts.

Surfaces and interfaces share some similarity to amorphous materials, where disruptions in bonding may lead to TLSs.
Extensive research efforts~\cite{de_leon_materials_2021,murray_material_2021} have lowered surface loss to a point where bulk dielectric loss in the crystalline substrates plays an important role in limiting coherence times~\cite{read_precision_2023}.
Indeed the dependence of loss on the device geometry and the distribution of fields within has been crucial in enabling the separation of surface and bulk loss components.
However, the microscopic source of dielectric loss in a bulk crystal, where bonding disruptions are infrequent, has not been re-assessed.

Here we examine the potential contribution of absorption due to magnetic dipole transitions to bulk loss in dielectric materials.
While such interactions are generally much weaker than those mediated by electric fields through electric dipole transitions, they can nonetheless contribute substantially to absorption~\cite{creedon_high_2011,dodson_magnetic_2012}.
In particular, we will focus here on defect-related transitions between levels that are split by zero-field splitting (ZFS), the lifting of the degeneracy of spin states in the absence of an applied magnetic field.
This splitting can be caused by the mutual interaction of unpaired electrons or by spin-orbit coupling when heavy elements are present.
Magnetic dipole-allowed transitions between the magnetic sublevels can be observed with electron paramagnetic resonance~\cite{lu_rapid_2017}.
Moreover the influence of spins on superconducting qubits has been noted~\cite{creedon_high_2011,schuster_high-cooperativity_2010,de_graaf_direct_2017,lee_identification_2014}, but to our knowledge the strength of the absorption of the electromagnetic field and the resulting loss has not previously been quantified.
Note that we use the term ``defect'' to refer to either native point defects or impurities in the host material.

In this work, we present a derivation of the absorption cross section for a magnetic dipole transition.
While the derivation follows standard references, we provide the context necessary to utilize the equations in the context of evaluating bulk loss.
In particular, we estimate the requisite parameters for the study of bulk loss in real materials.
As a case study, we apply the formula to calculate the absorption arising from the presence of Cr, Fe, and V impurities in sapphire.
This is a technologically relevant example because superconducting qubits utilize sapphire substrates, and it has been suggested that bulk loss in the substrate limits the coherence time of these qubits~\cite{read_precision_2023}.
While we focus on sapphire, the equations apply equally well to other relevant materials.
We find that bulk loss tangents for the ZFS absorption due to these transition-metal impurities are comparable to the loss measured experimentally, indicating that magnetic loss may play a role in limiting the coherence time of superconducting qubits.
We also more broadly discuss the expected magnitude of absorption due to ZFS\@.

\section{Results}
\begin{figure}[htb!]
    \centering
    \includegraphics[width=\columnwidth,height=0.6\textheight,keepaspectratio]{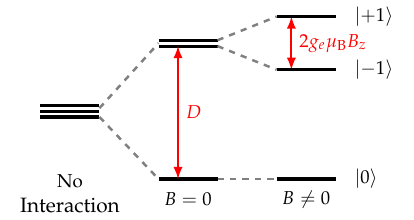}
    \caption{\label{fig:zfs}
        The magnetic sublevels of a $S = 1$ spin system with $C_{3v}$ symmetry.
        In the absence of interactions, the sublevels are degenerate.
        Even without a magnetic field, zero-field splitting separates the $\pm 1$ sublevels from the $0$ sublevel with a magnitude $D$.
        Applying a magnetic field along the high symmetry axis (defined as $z$) results in Zeeman splitting proportional to the magnitude of the magnetic field.
    }
\end{figure}

The prototype example for ZFS is the spin triplet, i.e., the $S$=1 spin system (Fig.~\ref{fig:zfs}).
A magnetic field causes Zeeman splitting, a separation of levels with different values of the magnetic spin quantum number $m_s$ ($-1$, 0, or +1).
However, even in the absence of a magnetic field, the triplet levels can be split due to spin-spin and spin-orbit interactions.
ZFS values can be as large as THz but are more typically in the GHz range.
For instance, the NV$^-$ center in diamond has a ZFS of 2.88 GHz~\cite{loubser_electron_1977}, and the value for a Cr$^{3+}$ ion in sapphire is 11.45 GHz~\cite{farr_ultrasensitive_2013,tobar_proposal_2003}.
While we focus on loss in superconducting qubits, it is worth noting that magnetic loss is an issue for a variety of other quantum platforms~\cite{de_leon_materials_2021}, for example in quantum defects~\cite{onizhuk_colloquium_2025}, and these equations apply equally well there.

To evaluate the absorption due to transitions between ZFS levels, we calculate the absorption coefficient $a (\omega)$ at photon frequency $\omega$.
The absorption coefficient is related to the attenuation of the electromagnetic radiation in the material by
\begin{equation}
    \label{eq:I}
    I(z) = I(0) \, e^{-a(\omega) z} \;,
\end{equation}
where $I$ is the intensity at coordinate $z$ along the direction of propagation.
The absorption coefficient can be related to the absorption cross section per defect $\sigma (\omega)$ by
\begin{equation}
    \label{eq:ktosigma}
    a(\omega) = N_{\rm def} \, \sigma(\omega) \;,
\end{equation}
where $N_{\rm def}$ is the defect density.

The absorption cross section for a magnetic dipole transition~\cite{stoneham_theory_1975} is derived in the Supplemental Material~\cite{sm} and given by:
\begin{equation}
    \label{eq:md}
    \sigma_{\rm MD} (\omega) = n_r \pi^2 \alpha^3 a_0^2 \, {\lvert {\bm \xi}_{\rm B} \cdot {\bf M}_{IF} \rvert}^2 \omega \, \delta (\omega - \Omega_{IF}) \;,
\end{equation}
where $n_r$ is the index of refraction, $\alpha$ is the fine-structure constant and $a_0$ is the Bohr radius.
The energy difference between the initial ($I$) and final ($F$) state is given by $\hbar\Omega_{IF}$.
${\bm \xi}_{\rm B} = \hat{\bf k} \times {\bm \xi}$, where ${\bm \xi}$ is the polarization vector and $\hat{\bf k}$ is the unit vector along the wavevector of the incoming light.
The transition matrix element ${\bf M}_{IF}$ is given by
\begin{equation}
    \label{eq:me}
    {\bf M}_{IF} =  \braket{\Psi_F \lvert \frac{{\bf L} + g_e {\bf S}}{\hbar} \rvert \Psi_I} \;,
\end{equation}
where ${\bf L}$ ({\bf S}) is the orbital (spin) angular momentum operator and $g_e$ is the electron $g$-factor.
$\Psi_{I/F}$ are the wavefunctions of the initial ($I$) or final ($F$) state.
The above definitions make ${\bf M}_{IF}$ dimensionless.

Experimentally it is common to measure the loss tangent $\tan(\delta)$, which is related to the absorption coefficient by
\begin{equation}
    \label{eq:loss_tan}
    \tan(\delta) = \frac{c}{n_r \omega} a(\omega) \; ,
\end{equation}
where $c$ is the speed of light.
The bulk loss tangent can be evaluated utilizing Eqs.~(\ref{eq:ktosigma})--(\ref{eq:me}).
Inspection of Eqs.~(\ref{eq:md}) and (\ref{eq:loss_tan}) shows that $n_r$ cancels out.

The delta function in Eq.~(\ref{eq:md}) is appropriate for a discrete transition;
however, in reality a variety of mechanisms broaden the transition.
At the basic level, energy-time uncertainty leads to homogeneous broadening due to the finite lifetime of states.
For homogeneous broadening, the delta function in Eq.~(\ref{eq:md}) should be replaced with a Lorentzian $L_\gamma (\omega)$:
\begin{equation}
    \label{eq:lorentzian}
    L_\gamma (\omega) = \frac{1}{\pi} \frac{\gamma / 2}{\omega^2 + (\gamma/2)^2} \;,
\end{equation}
where $\gamma^{-1}$ is the lifetime of the excited state.

Inhomogeneous broadening can also be present, where each defect feels a different local environment due to random fields (strain, electromagnetic, etc.).
Inhomogeneous broadening is typically modelled by replacing the delta function with a Gaussian function.
When both homogeneous and inhomogeneous broadening are present, a Voigt function (a convolution of the Lorentzian and Gaussian) can be used.
However, the tails of the Gaussian function fall off more rapidly than the Lorentzian.
As the strain dependence of ZFS is typically small~\cite{*[{See the Supplemental Information of }][{}] whiteley_spinphonon_2019}, the Lorentzian will dominate when far off resonance.
For the present case, we will consider only homogeneous broadening as the dominant broadening mechanism.

The above equations provide the necessary context to the textbook information on magnetic dipole transitions (given in the Supplemental Material~\cite{sm}) to apply them to study bulk loss.
As a case study for the calculation of the bulk loss tangent, we consider a transition between magnetic spin sublevels of a Cr impurity in sapphire.
In the dominant $3+$ oxidation state, Cr has a ground-state spin $S$ of 3/2.
We take the ZFS parameter $D$ and the electron $g$-factor $g_e$ from experiment~\cite{farr_ultrasensitive_2013,tobar_proposal_2003}.
$D$ for Cr is negative, indicating that the $m_s = \pm 3/2$ states are lower in energy than the $m_s = \pm 1/2$ states, and the separation between the two pairs $\Omega_{IF}/(2\pi)$ is 11.45~GHz.
Values for the frequency $\Omega_{IF}/(2\pi)$ and for  $g_e$ are listed in Table~\ref{tab:TMparameters}.
Spin-lattice coupling reduces the lifetime $\gamma^{-1}$ of these states to a value of 37~ns ($\gamma = 27$~MHz)~\cite{farr_ultrasensitive_2013,tobar_proposal_2003}.

\begin{table}
    \caption{\label{tab:TMparameters}
        The ground-state spin $S$, electron g-factor $g_e$, and transition energy $\Omega_{IF}$ for the transition-metal impurities considered in this work.
        Vanadium has multiple transitions due to hyperfine coupling to its nuclear spin.
        Concentrations $N_{\rm def}$ are typical for HEMEX sapphire and taken from Ref.~\onlinecite{alexandrovski_absorption_2001}.
    }

    \begin{ruledtabular}
        \begin{tabular}{ccccc}
            Impurity & $S$ & $g_e$ & $\Omega_{IF}/(2\pi)$ [GHz] & $N_{\rm def}$ [cm$^{-3}$] \\
            \hline
            Cr & 3/2 & 1.984 & 11.45 & $10^{17}$ \\
            Fe & 5/2 & 2.02 & 12.03 & $10^{17}$ \\
            V  & 3/2 & 2.029 & 8.68 & $10^{16}$ \\
            && 2.045 & 8.83 & \\
            && 2.055 & 9.02  &\\
            && 2.057 & 9.25  &\\
            && 2.052 & 9.49  &\\
            && 2.035 & 9.78  &\\
            && 2.017 & 10.08 & \\
            && 1.994 & 10.40 & \\
        \end{tabular}
    \end{ruledtabular}
\end{table}

The key ingredient necessary to evaluate the bulk loss tangent is the matrix element ${\bf M}_{IF}$ (Eq.~\ref{eq:me}).
Cr is a moderately heavy ion and exhibits some degree of spin-orbit coupling.
Both spin-spin and spin-orbit coupling contribute to the quoted $D = -5.723$~GHz ZFS value.
However, the spin-orbit coupling is weak enough that spin is still a good quantum number, and one can identify transitions between spin magnetic sublevels.
In the derivation of the transition from $m_s = \pm 3/2$ to $m_s = \pm 1/2$, we can assume the orbital wavefunction does not change (to a good approximation), only the spin.
As a result, $\braket{\Psi_F \lvert {\bm L} \rvert \Psi_I} = 0$.
We then have $\lvert \braket{{\pm 1/2} \lvert S_{x/y} \rvert {\pm 3/2}} \rvert = \hbar \sqrt{3} / 2$ and the $S_z$ component is zero.
In the absence of information on the crystal orientation or light polarization, we take the light to be unpolarized and obtain ${\lvert {\bm \xi}_{\rm B} \cdot {\bf M}_{IF} \rvert} = g_e/\sqrt{2}$.

\begin{figure}[htb!]
    \centering
    \includegraphics[width=\columnwidth,height=0.6\textheight,keepaspectratio]{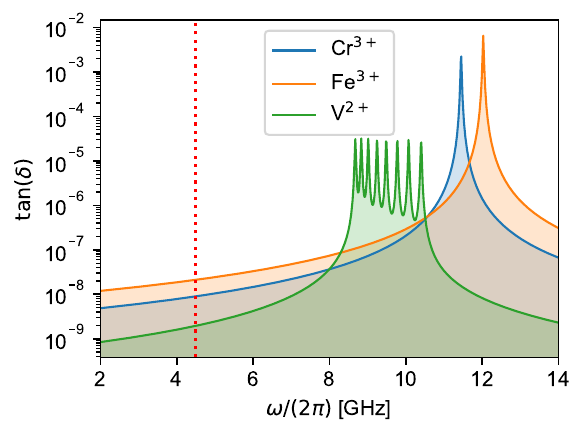}
    \caption{\label{fig:loss_tan}
        Calculated bulk loss tangents $\tan(\delta)$ resulting from magnetic dipole transitions at Cr (blue), Fe (orange) and V (green) impurities in sapphire.
        Superconducting qubits operate around $\omega/(2\pi) = 4.5$~GHz, shown by the vertical dashed line.
    }
\end{figure}

In order to obtain an absorption coefficient [Eq.~(\ref{eq:ktosigma})] and bulk loss tangent [Eq.~(\ref{eq:loss_tan})] we need to specify an impurity concentration $N_{\rm def}$.
Even the highest quality sapphire grown by the heat-exchanger method (HEMEX) has been found to contain Cr, Fe, and V impurities with concentrations $N_{\rm def}$ in the range $10^{16}$--$10^{17}$~cm$^{-3}$~\cite{alexandrovski_absorption_2001,route_heat-treatment_2004,prot_self-diffusion_1996,creedon_high_2011,read_dielectric_2024}; reported concentrations are listed in Table~\ref{tab:TMparameters}.
Using $N_{\rm def}$=$10^{17}$~cm$^{-3}$ for Cr, we calculate the bulk loss tangent $\tan(\delta)$ [Eq.~(\ref{eq:loss_tan})]
shown in Fig.~\ref{fig:loss_tan}.
On resonance with the transition, the bulk loss tangent can be as high as $\approx 10^{-3}$.
At $\omega/(2\pi) = 4.5$~GHz, the frequency relevant for qubits, we have $\tan(\delta) = 9.0 \times 10^{-9}$.

Cr is not the only paramagnetic impurity that is unintentionally present in sapphire~\cite{alexandrovski_absorption_2001,route_heat-treatment_2004}.
The measured concentrations, ground-state spins, and ZFS splittings for Fe and V are included in Table~\ref{tab:TMparameters}.
The resulting bulk loss tangents for Fe and V are included in Fig.~\ref{fig:loss_tan}.
While the dominant isotopes of Cr and Fe have zero nuclear spin, the dominant nuclear isotope of V has a spin of $7/2$, which leads to a hyperfine splitting on top of the ZFS\@.
As a result, the lowest frequency magnetic dipole transition occurs at 8.68~GHz.

\section{Discussion}
We now compare these results with experimental observations in the literature, but first we emphasize the distinction between bulk and surface loss.
Loss in typical superconducting qubits exhibits a dependence on device geometry and the distribution of fields within.
Modeling of this geometry dependence allows researchers to distinguish between surface and bulk contributions to loss~\cite{crowley_disentangling_2023,murray_material_2021}.
While surface contributions have long been dominant, bulk loss has recently come under scrutiny, and this is the main focus of our derivation.
Surface loss is, in our opinion, less mysterious than bulk loss: surfaces and interfaces share similarities to the amorphous materials that originated the concept of the TLS\@.
Of course, surface loss cannot be neglected.
Reductions in bulk loss renders surface loss more relevant, and vice versa.

Read \textit{et al.}~\cite{read_precision_2023} performed precision measurements of the microwave dielectric loss of bulk sapphire with parts-per-billion sensitivity using a specially designed ``dielectric dipper'' apparatus.
For HEMEX sapphire at 4.5 GHz they found a bulk loss tangent of $19 \times 10^{-9}$~\cite{read_precision_2023}.
Read \textit{et al.} bounded the magnetic loss to be less than 3\% in sapphire obtained by edge-defined film-fed growth, which had higher loss than in the higher quality HEMEX sapphire.
Our results demonstrate that, for realistic assumptions on HEMEX sapphire, the magnetic loss is comparable in magnitude to the loss in the electric field, and hence should not be overlooked as an important contribution to the overall loss.

Crowley \textit{et al.}~\cite{crowley_disentangling_2023} also reported results relating to bulk loss in HEMEX sapphire, but measured on superconducting resonators.
These devices showed bulk loss about an order of magnitude higher than the loss measured by Read \textit{et al.}~\cite{read_precision_2023}.
Crowley \textit{et al.}~\cite{crowley_disentangling_2023} separated the loss into two components and attributed one component to two-level systems.
The remaining loss (referred to as ``other'' in Ref.~\onlinecite{crowley_disentangling_2023}) was found to be comparable to that of Read \textit{et al.}~\cite{read_precision_2023}.
Crowley \textit{et al.}~\cite{crowley_disentangling_2023} hypothesized that the TLS-type loss was related to extended defects introduced during processing.
Based on our results we suggest that magnetic loss may also contribute.

Crowley \textit{et al.}~\cite{crowley_disentangling_2023} identified the TLS contribution based on temperature ($T$) and power ($P$) dependence, which is a common approach in device characterization to distinguish between different contributions to loss.
The ZFS mechanism discussed here effectively also acts as a TLS, and the $T$ dependence can be included in Eq.~(\ref{eq:md}) with a factor
\begin{align}
    \label{eq:Tdep}
    w(T) &= p_I (1-p_F)
    = (1 + e^{-\hbar\Omega_{IF} / k_{\rm B} T})^{-2} \;,
\end{align}
where $p_{I/F}$ is the phonon occupation factor and $k_{\rm B}$ is the Boltzmann constant.
The loss is independent of $T$ at low temperatures and saturates to a constant at high $T$.

Experimentally, the temperature dependence of TLSs is usually fit with a $\tanh(\hbar\Omega_{IF}/2k_{\rm B}T)$ function~\cite{gao_physics_2008}.
This function accounts for the behavior of \textit{resonant} TLSs:
at resonance, stimulated emission occurs, i.e., the field causes a new photon with frequency $\Omega_{IF}$ to be produced, thus compensating for the absorption of photons at that frequency.
Neglecting stimulated emission would be more appropriate off resonance, as we do here.
The hyperbolic tangent goes to zero at high $T$, whereas Eq.~(\ref{eq:Tdep}) approaches a non-zero constant.
In practice, a non-zero constant is always included in the fitting procedure, so that both types of fits will equally well model the high-$T$ behavior.
Therefore temperature dependence does not provide a way to distinguish between resonant and off-resonant contributions.

The power dependence can be addressed by having a $P$-dependent broadening~\cite{gao_physics_2008,hunklinger_ultrasonic_1976} in Eq.~(\ref{eq:lorentzian}):
\begin{equation}
    \label{eq:Pdep}
    \gamma (P) = \gamma \sqrt{1 + P/P_c} \;,
\end{equation}
where $P_c$ is the critical power.
On resonance, the loss decreases at higher powers, typically referred to as saturation.
Off resonance, the loss {\it increases} at higher powers.
Experimentally, saturation at high power is typically interpreted as evidence for TLSs, which would at face value exclude these paramagnetic impurities as relevant TLSs.
Characterization of TLSs has always revealed the presence of an ensemble of centers resonating at different frequencies.
This suggests that loss at a given frequency could be dominated by {\it off-resonance} contributions---which do not exhibit power saturation.
As discussed below, it may be worth re-examining whether power saturation is sufficient evidence of resonant TLSs.

Identifying the microscopic origin of TLSs has been an outstanding issue for many years~\cite{muller_towards_2019}.
TLSs are often associated with sharp resonances, a notion that arises from observations of avoided crossings resulting from coupling to individual TLSs in frequency-sweep experiments~\cite{martinis_decoherence_2005,simmonds_decoherence_2004,oh_elimination_2006}.
However, such experiments do not quantify the loss associated with an individual TLS\@.
As shown in Fig.~\ref{fig:loss_tan}, when considered as an ensemble, the TLSs arising from ZFS states of paramagnetic impurities can lead to significant loss at frequencies well away from the actual resonance.
We suggest that this may be an important effect on qubit coherence.
Given the comparable magnitude of the estimated loss with experiments,
these results beg the question whether the power saturation observed in experiments is actually due to resonant TLSs.
It is plausible that there is some mechanism that, in conjunction with off-resonant TLSs, explains the saturation.
For example, there could be cavity effects~\cite{karmstrand_unconventional_2023}, or local heating of the off-resonant TLSs~\cite{mittal_annealing_2024} could lead to an effective power saturation.

Finally, we comment more generally on the expected magnitude of absorption associated with magnetic dipole transitions.
Such transitions are frequently discussed in the context of emission from rare-earth ions, in which electric dipole transitions are often forbidden or very weak~\cite{dodson_magnetic_2012}.
Results reported for these transitions can provide information about the magnitude and trends of the matrix elements.
Examination of those treatments also yields insight into the case when spin-orbit coupling is strong and spin is no longer a good quantum number.

Dodson and Zia~\cite{dodson_magnetic_2012} evaluated a range of magnetic dipole transitions in rare-earth ions and tabulated those with the highest emission rates ($> 5$~Hz);
a selected subset are listed in Table~\ref{tab:rareearth}.
To extract the matrix elements from their spontaneous emission rates, we derive an expression for the rate in the Supplemental Material~\cite{sm}.
The resulting values for the matrix element $\lvert M_{IF} \rvert$ are listed in Table~\ref{tab:rareearth}.
The matrix elements for the rare-earth ions vary by more than a factor of 30, and thus we would expect the rates to vary by three orders of magnitude;
however, the actual emission rates are all within a factor of five.
This is due to the $\Omega_{IF}^3$ scaling of the emission rate compensating for the differences in the matrix elements.

\begin{table}
    \caption{\label{tab:rareearth}
        The wavelength $\lambda$, frequency $\Omega_{IF}$, emission rate $A_{MD}$, and our extracted matrix element $\lvert M_{IF} \rvert$ [Eq.~(S21) in the Supplemental Material~\cite{sm}], for selected transitions from Ref.~\onlinecite{dodson_magnetic_2012}.
        For comparison, we also include the values for Cr and Fe, as well as a representative value for V.
    }
    \begin{ruledtabular}
        \begin{tabular}{c c c c c}
            Element & $\lambda$ [nm] & $\Omega_{IF}/(2\pi)$ [THz] & $A_{MD}$ [Hz] & $\lvert M_{IF} \rvert$ \\
            \hline
            Gd & 307 & 977 & 30.24 & 0.0108 \\
            Sm & 477 & 628 & 7.14 & 0.0096 \\
            Eu & 700 & 428 & 24.63 & 0.1044 \\
            Er & 1276 & 235 & 12.21 & 0.3135 \\
            Dy & 1550 & 193 & 6.21 & 0.2858 \\
            Cr & $2.62 \times 10^7$ & 0.01145 & - & 1.403 \\
            Fe & $ 2.49 \times 10^7$ & 0.01203 & - & 2.332 \\
            V & $ 3.45 \times 10^7$ & 0.00868 & - & 1.435 \\
        \end{tabular}
    \end{ruledtabular}
\end{table}

In the case of a pure spin transition between two spin magnetic sublevels (i.e., without any change in the orbital configuration) we found the matrix element to be of order 1.
The matrix elements for the rare-earth transitions are smaller in magnitude, which we attribute to the mixing of states from spin-orbit coupling,
Indeed, when spin-orbit coupling is strong, a given initial or final state $\Psi_{I/F}$ is a linear combination of several states with the same total angular momentum.
States with the same total angular momentum can vary in the orbital and spin angular momentum.
This can lead to a reduction of the magnitude of the matrix elements compared to a pure spin transition, as well as the variation in the size of the matrix elements observed in Table~\ref{tab:rareearth}.

These insights into the magnitude of magnetic dipole matrix elements are based on transitions in the THz range of frequencies (Table~\ref{tab:rareearth}).
One may ask why we do not compare to spontaneous emission rates at GHz (microwave) frequencies.
The answer is that no such data are available.
This is related to the unfavorable $\Omega_{IF}^3$ scaling of the spontaneous emission rate, which indicates that a $\sim$GHz transition emits at a rate nine orders of magnitude slower than a $\sim$THz transition (roughly one photon per year), rendering it unobservable.
On the other hand, only \textit{one} factor of $\Omega_{IF}$ appears in the expression for the absorption cross section, explaining why absorption can still be observed at microwave frequencies.

\section{Conclusions}
In summary, we have derived the absorption cross section for a magnetic dipole transition with a focus on ZFS of paramagnetic impurities.
We applied the formula to the case of transition-metal impurities in sapphire and calculated the bulk loss tangents resulting from the absorption.
Our results demonstrate that the magnetic loss can be comparable to the loss in the electric field that has been observed experimentally.
We also suggest that magnetic dipole transitions related to ZFS can be an origin of TLSs that have been found to limit the coherence times of superconducting qubits.

\begin{acknowledgements}
    We gratefully acknowledge fruitful discussions with A. P. Read, N. P. de Leon, and R. J. Schoelkopf.
    This work was supported by the U.S. Department of Energy, Office of Science, National Quantum Information Science Research Centers, Co-design Center for Quantum Advantage (C2QA) under contract number DE-SC0012704.
    The research used resources of the National Energy Research Scientific Computing Center, a DOE Office of Science User Facility supported by the Office of Science of the U.S. Department of Energy under Contract No. DE-AC02-05CH11231 using NERSC award BES-ERCAP0021021.
\end{acknowledgements}

\section*{Data Availability}
The data that supports the findings of this study can be obtained readily from the figures or are available from the corresponding author upon reasonable request.

\end{document}


\title{Supplemental Material for\\Impact of Absorption due to Zero-Field Splitting on Loss in Dielectrics}

\author{Mark E. Turiansky}
\email{mturiansky@physics.ucsb.edu}
\altaffiliation[Present Address: ]{US Naval Research Laboratory, 4555 Overlook Avenue SW, Washington, DC 20375, USA}
\affiliation{Materials Department, University of California, Santa Barbara, CA 93106-5050, U.S.A.}

\author{Chris G. Van de Walle}
\email{vandewalle@mrl.ucsb.edu}
\affiliation{Materials Department, University of California, Santa Barbara, CA 93106-5050, U.S.A.}

\date{\today}

\maketitle

\section{Derivation of the Absorption Cross Section}
\label{sec:deriv}

Following the derivation in Chapter 10.1 of Ref.~\onlinecite{stoneham_theory_1975} we calculate the absorption cross section $\sigma$:
\begin{equation}
    \label{eq:acs0}
    \sigma (\omega) = \frac{r}{S / \hbar \omega} \;,
\end{equation}
where $r$ is the rate of the absorption process.
$S$ is the magnitude of the time-averaged Poynting vector, which quantifies the energy flux and has units of J~m$^{-2}$~s$^{-1}$.
$S / \hbar \omega$ is thus the photon flux.

Based on Fermi's golden rule, we can write the transition rate $r$ as
\begin{equation}
    \label{eq:r}
    r = \frac{2\pi}{\hbar^2} {\lvert \braket{\Psi_F \lvert H^\prime \rvert \Psi_I} \rvert}^2 \delta (\omega - \Omega_{IF}) \;,
\end{equation}
where $H^\prime$ is the perturbation, $\Psi_{I}$ ($\Psi_{F}$) is the wavefunction of the initial $I$ (final $F$) state, and $\hbar\Omega_{IF}$ is the energy difference between the states.

We consider the electromagnetic radiation defined by the vector potential ${\bf A}$:
\begin{equation}
    \label{eq:A}
    {\bf A} ({\bf r}, t) = {\bm \xi} \, A_0 \cos ({\bf k} \cdot {\bf r} - \omega t) \;,
\end{equation}
where ${\bm \xi}$ is the polarization vector and ${\bf k}$ the wavevector.
The wavevector is related to the frequency of the radiation by $k = n_r \omega / c$, where $n_r$ is the index of refraction.
The energy flux $S$ is then given by
\begin{equation}
    \label{eq:S}
    S = \frac{n_r \omega^2 A_0^2}{2 \mu_0 c} \;.
\end{equation}
Thus the absorption cross section is
\begin{equation}
    \label{eq:acs1}
    \sigma (\omega) = \frac{4 \pi}{n_r c \epsilon_0 \hbar \omega A_0^2} {\lvert \braket{\Psi_F \lvert H^\prime \rvert \Psi_I} \rvert}^2 \delta (\omega - \Omega_{IF}) \;.
\end{equation}

The perturbation $H^\prime$ is obtained from minimal coupling and is given by
\begin{equation}
    \label{eq:H}
    H^\prime = -\frac{e A_0}{2 m_e} e^{i {\bf k} \cdot {\bf r}} \, {\bm \xi} \cdot {\bf p} \;,
\end{equation}
where ${\bf p}$ is the electron momentum.
For electromagnetic radiation at microwave frequencies, ${\bf k} \cdot {\bf r}$ is a small quantity, and we can perform a Taylor expansion:
\begin{equation}
    \label{eq:Ht}
    H^\prime = -\frac{e A_0}{2 m_e} \left( 1 + i {\bf k} \cdot {\bf r} + \dots \right) {\bm \xi} \cdot {\bf p} \;.
\end{equation}
The first term in Eq.~(\ref{eq:Ht}) corresponds to the electric dipole transition:
\begin{equation}
    \label{eq:HED}
    H^\prime_{\rm ED} = -\frac{e A_0}{2 m_e} {\bm \xi} \cdot {\bf p} \;. 
\end{equation}
Inserting Eq.~(\ref{eq:HED}) into Eq.~(\ref{eq:acs1}) and using
\begin{equation}
    \label{eq:ptor}
    \braket{\Psi_F \lvert {\bf p} \rvert \Psi_I} = i \Omega_{IF} m_e \braket{\Psi_F \lvert {\bf r} \vert \Psi_I} \; 
\end{equation}
we obtain
\begin{equation}
    \label{eq:sigma_ed_pre}
    \sigma_{\rm ED} (\omega) = \frac{4 \pi^2 \alpha}{n_r} \frac{\Omega_{IF}^2}{\omega} {\lvert {\bm \xi} \cdot {\bf r}_{IF} \rvert}^2 \delta (\omega - \Omega_{IF}) \;,
\end{equation}
where $\alpha$ is the fine-structure constant and ${\bf r}_{IF} = \braket{\Psi_F \lvert {\bf r} \rvert \Psi_I}$ is the dipole matrix element.

Following the common practice of replacing $\omega$ in the denominator of Eq.~(\ref{eq:sigma_ed_pre}) with $\Omega_{IF}$ (because of the delta function),
we thus obtain the absorption cross section for an electric dipole transition:
\begin{equation}
    \label{eq:sigma_ed}
    \sigma_{\rm ED} (\omega) = \frac{4 \pi^2 \alpha}{n_r} {\lvert {\bm \xi} \cdot {\bf r}_{IF} \rvert}^2 \Omega_{IF} \, \delta (\omega - \Omega_{IF}) \;.
\end{equation}

We examine the second term in Eq.~(\ref{eq:Ht}) by focusing on the case of ${\bf k} \sim \hat{z}$ and ${\bm \xi} \sim \hat{y}$ and rewriting the expression as follows:
\begin{align}
    \label{eq:second_term}
    ({\bf k} \cdot {\bf r}) ({\bm \xi} \cdot {\bf p}) &= \frac{n_r \omega}{c} i\hbar z \frac{\partial}{\partial y} \nonumber \\
    & = \frac{n_r \omega}{c} \frac{i\hbar}{2} \left( z \frac{\partial}{\partial y} + z \frac{\partial}{\partial y} \right) \nonumber \\
    &= \frac{n_r \omega}{c} \frac{i\hbar}{2} \left( z \frac{\partial}{\partial y} - y \frac{\partial}{\partial z} + y \frac{\partial}{\partial z} + z \frac{\partial}{\partial y} \right) \nonumber \\
    &= \frac{n_r \omega}{c} \frac{i\hbar}{2} \left( L_x / \hbar + y \frac{\partial}{\partial z} + z \frac{\partial}{\partial y} \right) \;,
\end{align}
where $L_x$ is the $x$ component of the angular momentum operator.
The first term in Eq.~(\ref{eq:second_term}) defines the magnetic dipole, while the remaining terms correspond to the electric quadrupole.
A full derivation starting from the Dirac equation would additionally result in coupling of the electron spin to the radiation:
the standard procedure is to replace $L_x \rightarrow L_x + g_e S_x$~\cite{stoneham_theory_1975}, where $g_e$ is the electron $g$-factor.
We can then define the perturbation for the magnetic dipole transition as
\begin{align}
    \label{eq:HMD}
    H^\prime_{\rm MD} &= \frac{e A_0}{2 m_e} \frac{\hbar \omega n_r}{2 c} (\hat{\bf k} \times {\bm \xi}) \cdot {\bf M} \nonumber \\
    &= \frac{e A_0 \omega n_r \alpha a_0}{4} (\hat{\bf k} \times {\bm \xi}) \cdot {\bf M} \;,
\end{align}
where $a_0$ is the Bohr radius and
\begin{equation}
    \label{eq:nu}
    {\bf M} = \frac{{\bf L} + g_e {\bf S}}{\hbar} \;,
\end{equation}
is the magnetic dipole operator, which we have defined to be dimensionless.
Inserting Eq.~(\ref{eq:HMD}) into Eq.~(\ref{eq:acs1}), we obtain
\begin{equation}
    \label{eq:sigma_md}
    \sigma_{\rm MD} (\omega) = n_r \pi^2 \alpha^3 a_0^2 \, {\lvert {\bm \xi}_{\rm B} \cdot {\bf M}_{IF} \rvert}^2 \omega \, \delta (\omega - \Omega_{IF}) \;,
\end{equation}
where ${\bf M}_{IF} = \braket{\Psi_F \lvert {\bf M} \rvert \Psi_I}$ and ${\bm \xi}_{\rm B} = \hat{\bf k} \times {\bm \xi}$.

\section{Spontaneous Emission rate of Magnetic Dipole Transitions}
\label{sec:mdse}

By invoking detailed balance~\cite{stoneham_theory_1975}, the spontaneous emission rate can be related to the absorption cross section $\sigma$ by
\begin{equation}
    \label{eq:spon_em_cross}
    A = \int d\omega \; \frac{c}{n_r} \, \sigma (\omega) \, \rho (\omega) \;,
\end{equation}
where $\rho$ is the photon density of states and photons travel at a velocity $c/n_r$ in the medium.
The density of states is the number of photons per frequency per unit volume,
\begin{equation}
    \label{eq:dos_def}
    \rho (\omega) = \frac{1}{V} \frac{dN}{d\omega} \;,
\end{equation}
where $N$ is the number of photons in a volume $V = L^3$.
Consider a shell of radius $k$, where $k$ is quantized.
The shell has a volume of $4 \pi k^2 dk$, and within that shell there are $4 \pi k^2 dk / (2\pi/L)^3$ photon states.
There are two equivalent polarizations, so we obtain
\begin{equation}
    dN = 2 \frac{4 \pi k^2 dk}{8\pi^3 / L^3} \;.
\end{equation}
Combining this with Eq.~(\ref{eq:dos_def}) and utilizing $k = n_r \omega / c$, we obtain
\begin{equation}
    \label{eq:dos}
    \rho(\omega) = \frac{n_r^3 \omega^2}{c^3 \pi^2} \;.
\end{equation}

The spontaneous emission rate for a magnetic dipole transition is then
\begin{align}
    A_{\rm MD} &= \frac{c}{n_r} \frac{n_r^3 \Omega_{IF}^2}{c^3 \pi^2} n_r \pi^2 \alpha^3 a_0^2 \, \Omega_{IF} \lvert M_{IF} \rvert^2 \nonumber \\
    \label{eq:A_MD}
    &= \frac{n_r^3 \alpha^3 a_0^2}{c^2} \Omega_{IF}^3 \lvert M_{IF} \rvert^2 \;,
\end{align}
where we have defined
\begin{equation}
    \label{eq:unpol_me}
    \lvert M_{IF} \rvert^2 = \frac{1}{3\hbar} \sum_\alpha {\lvert \braket{\Psi_F \lvert L_\alpha + g_e S_\alpha \rvert \Psi_I} \rvert}^2 \;,
\end{equation}
for unpolarized light.

\bibliographystyle{apsrev4-2}